# Title: Policy Robustness & Uncertainty in Model-based Decision Support for the Energy Transition

**Authors**: Ian J. Burton, Femke J.M.M. Nijsse, James M. Salter

**Corresponding emails:** ib400@exeter.ac.uk, f.j.m.m.nijsse@exeter.ac.uk, j.m.salter@exeter.ac.uk

**Keywords**: uncertainty quantification, energy modelling, uncertainty analysis, power systems, emulation, climate policy, policy analysis, decision-making

**Abstract**

Climate policy modelling is a key tool for assessing mitigation strategies in complex systems, where uncertainty is inherent and unavoidable. We present a general methodology for extensive uncertainty analysis in this field. While other studies have performed uncertainty analyses, few apply methods from the field of Uncertainty Quantification, which are commonly used in other modelling disciplines. We show how emulators can identify key uncertainties in modelling frameworks and demonstrate a novel policy analysis previously restricted by computational cost and limited representation of uncertainty. We apply this methodology to FTT:Power to explore uncertainties in the electricity system transition both globally and in India to assess the robustness of mitigation strategies to a wide range of policy and techno-economic scenarios. This approach results in much larger uncertainties in transition outcomes than commonly represented, but policy design can be shaped to mitigate this. Globally, our results indicate transition uncertainty is dominated by average rates of renewables cannibalisation, construction times and grid connection lead times, outweighing regional price policies, including policy reversals in the US. Solar PV appears most resilient due to low costs, though still sensitive to infrastructure constraints and cannibalisation. Onshore wind is more exposed to a range of uncertainties. In India, we find evidence that policy packages including partial phase-out instruments have greater robustness to key uncertainties, although longer lead times still hinder policy goals. Our results suggest that enabling policy and regulating fossil fuels are critical for robust power sector transitions.

## 1. Introduction

Making policies in complex and unpredictable domains is challenging, and analysts often use computational models. These show how policy options function within the model and provide insight into their possible effects in reality (Kwakkel et al. 2010; Walker et al. 2003). However, as the relationship between models and reality is not fully known, uncertainty is inevitable. Some policies may work well in particular model settings but lose their effectiveness under other plausible conditions. In complex policy environments, such as climate or energy policy, uncertainty is pervasive and there is need for more in-depth analysis that goes beyond assessing effectiveness under a narrow set of assumptions (Saltelli et al. 2020).

Designing policy that remains effective under uncertainty—so-called robust policy—has emerged as a key goal in policy studies (Dewulf & Biesbroek 2018). More specifically, robustness is defined as a policy property that empowers it to retain its functionality under uncertainty, to fulfil its intended purposes and objectives across possible future contexts (Capano & Woo 2018; Howlett et al. 2018).

Robustness differs from resilience which emphasises 'returning to a stable equilibrium point after a shock' (Capano & Woo 2018; 2017) rather than the ability to attain goals under shocks and uncertainty. A distinction is also made between robustness of policy 'output' and 'process'. Output refers to the policies themselves, and policy process refers to the wider policy formulation and implementation cycle such as institutional decision-making frameworks. For example, systemic risk analysis or decision-making under deep uncertainty (DMDU) approaches provide systematic frameworks to guide policymaking across both dimensions (Gambhir et al. 2025; Marchau et al. 2019). Policy processes often include structural and institutional features that, whilst crucial, are typically beyond the remit of policy analysts (Kwakkel et al. 2010). This paper focuses on the 'output' aspect of policy robustness from a modelling perspective. Studies have examined the qualitative characteristics of robust policies (Kwakkel et al. 2016), however, these are less often complemented by computational modelling tools. Uncertainty analysis often features in climate policy modelling but the direct application of such analysis for designing robust policies is rare prompting calls for greater treatment of uncertainty in the field (Workman et al. 2021; Panos et al. 2023).

Several qualitative frameworks have been developed that categorise uncertainty (Kwakkel et al. 2010; Stirling 2010) but from a modelling perspective, these uncertainties are broadly classed as 'parametric uncertainty', focused on input values, or 'structural' uncertainty (Pastor et al. 2020). Structural uncertainty refers to the discrepancy between models' dynamics and relationships and those of the system of interest and is typically investigated by multi-model comparisons (Iyer et al. 2015). However, structural differences between models can make them hard to compare. In energy system modelling, for example, key drivers of investment decisions between generation technologies are modelled in different ways and at different scales. Profit-maximisation by investors is endogenously modelled in Generation Expansion Planning frameworks (Brown & Reichenberg 2021) and equlibrium models of electricity markets (Ehrenmann & Smeers 2011). These detailed, granular approaches come with significant computational burden which makes capturing global dynamics more difficult (Gong et al. 2023). Some approaches abstract the investment decision process using levelized-cost comparisons and capture global feedbacks such as learning-by-doing (Mercure 2012). Direct comparisons between such different approaches are difficult but important in contextualising any modelling research.

Parametric uncertainty concerns the values of a model's parameters and can be analysed readily, primarily through scenario or sensitivity analysis (Pastor et al. 2020). Scenario analysis assigns specific values to parameters to simulate different system states (Brugnach et al. 2008), sensitivity analysis (SA) supplements this by estimating the influence of variation in those parameters on model outputs (IEA 2024). Common approaches rely on many simulations of the model, which may have high computational costs (Chiani et al. 2025). Local SA focuses on a subset of input space and is common in climate policy modelling studies but often criticised for providing limited insight (Razavi et al. 2021). Global SA attempts to represent the full input space (Saltelli 2008). Chiani et al.(2025) detail multiple approaches for global SA based on Monte Carlo methods, and propose their own strategies for reducing computational cost but acknowledge remaining constraints (Gamboa et al. 2014; Strong et al. 2015).

Scenario and sensitivity analysis are the primary tools for analysing uncertainty in climate policy modelling, such as for power sector transitions, and mainly focus on parametric uncertainty.

The International Energy Agency (IEA) models several global scenarios of ambition in decarbonisation of the energy sector. In their modelling, they find that rapid renewables deployment could put pressure on supporting activity such as permitting or supply chain. They project that emissions will fall nearly 30% by 2035 and approximately 50% by 2050 in their 2024 analysis on stated policies (STEPS, IEA 2024). They find uncertainty in electricity demand is driven by electrification in other industries, unpredictable weather patterns, and the rapid growth of new sectors such as AI data centres. In the accompanying SA, they explore the impact of variation in certain key factors, such as EV sales, electricity demand, fuel use and subsequent $CO_2$ emissions. This analysis suggests 2030 emissions could be 1.6% higher or lower than projected due to key uncertainties. More ambitious scenarios are also projected with varying assumptions about renewables deployment, coal phase-outs, energy storage support and demand flexibility. Other analyses identify other sources of risk, such as the tendency of higher shares of renewables to result in lower market values for renewables in liberalised energy markets. This 'cannibalisation' raises concerns over reduced returns on investment and its implications for system transitions (Hirth 2013).

Similar analyses are also applied to regional transitions. India is the third largest energy consumer with a power system that faces many uncertainties such as rapidly rising demand with electrification in multiple sectors, a growing population, and limited domestic gas reserves (Vijayalaxmi & Manthanwar 2024; Bhatia 2023). Studies examining key risks affecting renewables uptake have found electricity demand, technology costs, supporting infrastructure development and storage capacities are key sources of uncertainty (Chaturvedi et al. 2021; Dasgupta & Sarangi 2021; Singh & Idrisi 2020). Policy strategies have included feed-in tariffs, renewable purchase obligations, tradable renewable energy certificates or transmission cost waivers (Sawhney 2021; Vijayalaxmi & Manthanwar 2024). Scenario projections show solar will continue to expand rapidly and may come to dominate (Singh & Idrisi 2020). The Central Electricity Authority (CEA) in India projects between 53–55% renewables energy share (of capacity) by 2030. These projections are dependent on demand levels, hydropower availability and the cost of batteries which are explored through SA (CEA 2023).

The field of Uncertainty Quantification (UQ) offers a means of addressing model uncertainty more comprehensively by replacing complex models with computationally efficient emulators that approximate the original model (O'Hagan 2006). This enables robust uncertainty analysis that would otherwise be constrained by computational cost (Lee et al. 2011). Developing climate policy that is robust to uncertainty in assumptions and model parameters requires tools that explicitly capture conditional uncertainty within modelling frameworks (Stirling 2010).

Emulators are not widely used in climate or energy policy modelling. For example, the modelling framework used in this paper, FTT, has adopted uncertainty analysis based around model runs, whether by varying the inputs systematically (SA, Lam et al. 2018) or randomly (Monte Carlo, Nijsse et al. 2023). There are many similarities between these model-run based approaches and emulator-based approaches, and scenario and sensitivity analyses can similarly be performed with emulators, but with a more flexible and rigorous approach possible due to their efficiency, opening up possibilities for policy analysis.

This study develops a general methodology, based on UQ, to identify key uncertainties in climate policy and to explore the robustness of policy outcomes. We provide a thorough analysis of model uncertainties at the global and regional level and use this to explore what policy packages show more robustness to key uncertainties for achieving certain policy goals. Our contribution advances the use of emulation in climate policy modelling and showcases an application to analyse policy robustness. In section 2, we describe our method and its application to the power system transition model, FTT:Power. We use this framework to analyse key uncertainties in the global electricity system transition and analyse policy robustness for India, with results presented in section 3. In section 4 we discuss our findings in relation to similar research, highlight our contribution and conclude.

## 2. Methodology

We explore uncertainties affecting power system transitions globally and regionally and investigate how different policy packages in India perform across a broad range of scenarios. By constructing emulators for model outputs, we explore future scenarios, associated uncertainties and the performance of policy packages under thousands of plausible techno-economic and political contexts. This allows us to assess policy robustness to these uncertainties, without requiring infeasible amounts of model simulations. Our study applies this general methodology to FTT:Power, but it could be applied to a broader set of models and questions.

Figure 1 displays the general workflow which has four broad parts. Parameterisation involves the identification of key inputs for outputs of interest, plausible parameter ranges and mapping from inputs to outputs. Simulations of the model under these parameterisations are performed to produce training data. Emulators are trained on this data and validated. Finally, the emulators are used to predict the model output under a wide range of scenarios for analysis.

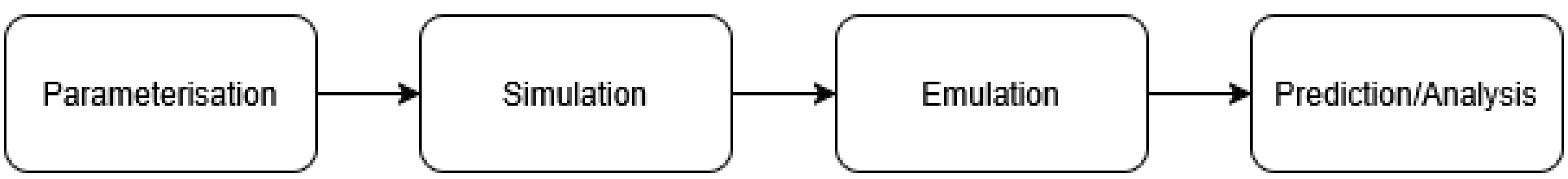

***Figure 1. Workflow****. The key stages of the methodology in sequential order moving from initial configuration through to the execution stages and finally to the analysis of the results.*

### 2.1 Parameterisation

#### 2.1.1 FTT:Power

The model family known as Future Technological Transformations (FTT) uses dynamic simulation modelling to understand how technology mixes in various sectors are set to develop in the coming decades (Mercure 2012). FTT:Power is a 71-region model of power sector investors and dispatchers and features 22 electricity generation and storage technologies. Residual load-duration curves are used to model load bands and storage needs; dispatchers decide on which generation asset to use by comparing marginal costs. The model is based on evolutionary dynamics simulating the uptake of technologies in S-curve processes characteristic of innovation. The central aspect is a set of replicator dynamics equations (Lotka-Volterra equations) which model the diffusion of technologies based on comparisons of levelized costs in chains of binary discrete choice models. The model incorporates a wide array of policy instruments, including regulations, taxes and subsidies. Innovation is incorporated by adding technology-specific cost reductions proportionate to cumulative global capacity. A term is added to costs to capture non-financial factors through calibration to historical shares data. This implicitly accounts for non-captured policies.

FTT: Power abstracts away from some key underlying dynamics that drive investment decisions in electricity markets. Whilst this is common in many global scale models, there are families of models, such as those mentioned above, which maintain greater granularity and better capture relevant dynamics. This has important implications for the analysis. FTT uses 'levelized-cost-of-electricity' (LCOE) calculations to compare how attractive different technologies are to investors by comparing costs and revenues over the entire lifetime of a technology. Dynamics affecting intermittent renewables such as greater need for storage, associated generation losses and curtailment are incorporated into the LCOE calculations. However, LCOEs have been criticised as misidentifying key drivers of investment decisions in liberalised energy markets. Rather, profit expectations are posited as the key driver which are explicitly modelled under mathematical frameworks mentioned above (Dagoumas & Koltsaklis 2019). For risk-averse, profit-maximising investors this is likely to make renewables less attractive (Ehrenmann & Smeers 2011) particularly in a context of limited long-term contracts (Dimanchev et al. 2024).

The model discrepancies resulting from abstractions in FTT need to be considered carefully when interpreting results, particularly given our goal is to better understand uncertainty and policy robustness to it. We refer to this in the discussion. Nevertheless, the alternative approaches mentioned come with significant computational burden (Dimanchev et al. 2024) and are usually employed at the national or regional level and certain dynamics are harder to capture. FTT is global and feeds into a macroeconomic model, E3ME, which makes it well-suited to exploring uncertainties and scenarios at this scale.

### 2.1.2 Energy Transition Outputs

We identify four key energy transition outputs in FTT which we train emulators for. These consist of electricity generation capacities for key renewables, namely solar photovoltaic and onshore wind, power sector emissions, shares of renewable capacity in the electricity mix and the weighted costs of electricity generation. To generate a set of FTT simulations for exploring the key uncertainties relating to these outputs and hence training emulators for them, we select a set of techno-economic and policy inputs to vary. We identify those that may be more influential based on relevant literature and discussions with model users. These inputs are detailed below.

### 2.1.3 Techno-economic Inputs

Techno-economic inputs determine key aspects of the energy system and the context within which policies function. Table 1 shows some of those varied here, their assumed ranges and whether these vary generally, regionally or by technology. We use parameter mappings that apply to multiple FTT:Power inputs, like learning rates for technologies which apply to all regions uniformly, or that scale pre-existing values making them relative and region-specific. For example, discount rates have a baseline value for each region-technology combination, and our single discount rate parameter acts on all these, assuming they vary in a correlated fashion. We make this simplifying assumption for flexibility and tractability.

| Input | Technologies / Regions | Ranges | Description/Sources |
|---|---|---|---|

| ***Learning Exponents*** | Solar PV | (-0.473) – ( -0.165)<br>Mean -0.319 | Key parameter for the learning rate (% price decline per doubling of capacity)<br>(Way et al. 2022; Krey et al. 2019) |
|---|---|---|---|
| | Onshore wind<br>Offshore wind | (-0.3) – (-0.088)<br>Mean -0.194 | |
| ***Lifetimes*** | Solar PV | 25–35 (years) | |
| | Onshore wind<br>Offshore wind | | |
| ***Lead times*** | Solar PV | 0.5-3 (years) | Lead times split into technology specific and system-wide timescales. Ranges devised from Gumber et al. (2024) |
| | Onshore wind | 1–3 (years) | |
| | Grid connection / commissioning | 0-1 (years) | |
| ***Discount Rates*** | Region x Technology | +/- 3% (2% minimum) | IRENA, BNEF |
| ***Electricity Demand*** | Regional | +/- 20% to growth rate of projected demand | E3ME results based on IEA projections |
| ***Fuel Prices*** | Coal<br>Gas (CCGT) | +/- 2 standard deviations from regional mean | Average over last 5 years (IEA) |
| ***Cannibalisation Rates*** | Solar PV | (-3.3) – (-0.2) | % decrease in technology value factor arising from 1% increase in market share. Qualitatively derived from (Stiewe et al. 2025; Hirth 2016, 2013)) |
| | Onshore<br>Offshore | (-2) – (-0.1) | |

***Table 1. Techno-economic input details.*** *Each input is listed with the technology/region combination it applies to within FTT:Power. Ranges provided as well as their sources.*

The learning exponent is the key parameter in standard equations for determining learning rates which represent the speed at which costs decline for a given technology with cumulative deployment. A lower learning exponent represents faster cost declines. This parameter has been heavily researched for many technologies as a way of understanding the differentiated cost declines from technologies such as fossil fuels or CCS, which have experienced very low cost declines, and technologies like solar photovoltaics which have experienced exponential cost declines (Stiewe et al. 2025; Hirth 2013a, 2016). Learning rates for similar technologies, e.g. onshore and offshore wind, are varied together under the assumption they are heavily correlated.

Average lifetimes for renewables vary in the literature which has significant implications for expected returns on investments especially given different preferences concerning weightings of future costs and revenues (Krey et al. 2019). Bringing a generation plant online involves multiple stages from first feasibility assessments through to when electricity actually starts to be produced (Gumber et al. 2024). How long this process takes, the lead time, is a key consideration when analysing a project's viability and determines how quickly investors can start to receive revenue and how quickly a technology can fill demand gaps from plant closure. Here, we define lead times as the period from construction start to electricity generation with grid connection delays treated separately.

Within FTT, the discount rate refers to the rate at which future revenues and costs are discounted when incorporated into the LCOE equations, reflecting the opportunity cost of capital and investor preferences for risk (Nasr et al. 2025). If the discount rate is higher, future costs and revenue have a smaller weighting in these comparisons making front-loaded high investment projects less attractive. Discount rates relate to interest rates as cheaper finance means lower discount rates, making long-term and riskier projects more viable. These preferences may vary depending on regional contexts and we have chosen baseline values on empirical data from BNEF and IRENA. We have varied these baseline values +/-3% given the spread of values that we observe in the data and set a lower limit of 2% as we do not observe any discount rates falling below this level. A similar approach is taken for electricity demand where regional baseline levels are taken from E3ME results which are based on IEA projections. The range of 20% either side of the growth rate in these projections was determined to be sufficient to cover a wide range of scenarios including ambitious electrification of sectors or even demand reductions. For fuel prices, we take the mean and standard deviations for each region over the last 5 years.

In liberalised energy markets, higher shares of renewables tend to reduce market value through 'cannibalisation', raising concerns about the implications for energy transitions from reduced returns on investment (Hirth 2013). Cannibalisation occurs when high renewable penetration depresses electricity prices during periods of abundant generation, reducing the market value of renewable output (Hirth 2013). Whilst this general dynamic is well established, there is little quantitative research beyond Europe and the US, limiting its applicability in global models such as FTT. To avoid overestimating certainty, we have qualitatively derived wide ranges for the rate of cannibalisation from an overview of the existing literature (Stiewe et al. 2025; Blume-Werry et al. 2021; Hirth 2013, 2016). We have increased the bounds in an effort to incorporate empirical uncertainty around rates in different contexts and the mitigating effects of flexibility mechanisms such as electricity storage, interconnectivity and demand-side responses (Hirth 2016).

#### 2.1.4 Policy Inputs

There are many ways policies could be parametrised in the model, FTT, which has specific input values for each year, instrument, and industry combination for 71 regions. To simplify the analysis and avoid unrealistic scenarios, we group certain inputs together and map newly constructed parameters onto these groups. The instruments analysed in this study are by no means comprehensive and the choices surrounding groupings and correlations are flexible. Regional and policy instrument inputs are mapped in different ways, generally guided by assumptions of what we expect to be correlated, such as subsidy support for both onshore and offshore wind power in the European Union. Groupings were also assessed for whether they allow for more meaningful understanding of uncertainty, such as the clustering of many individual countries into Global South and Global North groups. We remain sensitive to the issues arising from these 'reductive-aggregative' choices (Stirling 2003). However, for this analysis, uncertainty at the individual country level is overshadowed by other sources of variation, so aggregating countries provides clearer insight into global dynamics

For each region-instrument combination, we define a single policy parameter that varies on [0,1] and maps onto these sets. Through combining these policy ambition levels in different ways, certain types of instruments can be switched off entirely or simulated at any rate up to their regional maximum, allowing flexibility in the composition of policy packages simulated.

Table 2 details the three groupings of policy instruments, applied to technologies according to what is most common, depending on aspects such as the maturity of the technology. Phase-outs are applied equally to fossil fuel-based technologies such as coal, gas and oil. Subsidies are applied to nuclear, biomass, pumped hydro, geothermal and carbon capture and storage (CCS) technologies, and are correlated with feed-in tariffs applied to solar PV, onshore and offshore wind. Feed-in tariffs are differentiated based on the relative technology costs, with wind receiving a larger tariff. Carbon pricing is applied to all emitting technologies. The configuration is designed to balance between comprehensively representing the main types of instruments in relation to the generation technologies they are commonly applied to. There are notable omissions, for example we analyse carbon pricing and not cap-and-trade mechanisms which are present in many countries. This is due to a combination of the policy options available within the model and the likely effect between the two types of policy which we expect to be similar.

For regional groupings, China, the United States (US) and India are treated individually while the rest of the world is categorised into Global South or Global North based on World Bank classifications, giving 15 policy parameters (5 regions, 3 policy groupings) controlling levels of policy ambition. For example, the phase-out parameter for China can vary from no phase-out (0) to total phase-out (1), with mid-level phase-out (0.5) resulting in half the capacity additions that would occur under no phase-out. One exception is the feed-in tariff for the US which can also capture rollbacks of policy supporting renewables. For the US, parameter values between 0.5-1 represent the instrument levels in Table 2, values between 0.5-0 represent baseline policy for all instruments except feed-in tariffs which become negative. This is designed to capture removal of tax credits under the Inflation Reduction Act (IRA) by making renewables less attractive. The more ambitious policy levels are additional to the baseline which implicitly captures current policy for feed-in tariffs, subsidies and phase-outs. The baseline policy includes the current carbon price for regions with carbon pricing; this can be raised for mid-level and high-level policy.

| Instrument | Baseline policy | Mid-level policy | High-level policy |
|---|---|---|---|
| **Parameter value (excl. US)** | **0** | **0.5** | **1** |
| ***Phase-outs (sales restricted)*** | | | |
| *Oil, CCGT, Coal* | 0% | 50% | 100% |
| ***Feed-in tariffs*** | | | |
| Onshore/Offshore | \$0/MWh | \$20/MWh | \$40/MWh |
| Solar PV | \$0/MWh | \$15/MWh | \$30/MWh |
| ***Subsidies (upfront cost reduction)*** | | | |
| *Nuclear* | 0% | 10% | 20% |
| *CCS (coal, gas, waste)* | 0% | 25% | 50% |
| *Biomass (+ CCS)* | 0% | 30% | 60% |
| *Geothermal* | 0% | 25% | 50% |
| *Pumped Hydro* | 0% | 25% | 50% |
| ***Carbon Pricing*** | | | |
| *Oil, CCGT, Coal* | Current regional levels | \$31/ton$CO_2$ (2022) – \$345/ton$CO_2$ (2050) | \$62/ton$CO_2$ (2022) – \$564/ton$CO_2$ (2050) |

***Table 2. Policy instrument levels.*** *Each row signifies what instruments are assumed to be correlated and are varied with the same parameter. Three levels are shown, but parameter values can vary between 0 and 1.*

2.2 Simulation

Overall, our analysis features 30 variable parameters, split between 15 techno-economic and 15 policy inputs. To produce training data for emulation, 500 runs of FTT:Power are performed at input sets generated by Latin Hypercube Sampling, ensuring comprehensive exploration of the 30-dimensional parameter space (Urban and Fricker 2010). This has better space filling properties than Monte Carlo based sampling and ensures the emulators are trained on data as representative of behaviour across all input space as possible. Moreover, additional runs can be performed if regions of parameter space are predicted poorly, with validation checks to assess this.

### 2.3 Emulation

The analysis uses emulators that are trained on input and output data from FTT:Power model runs at varied inputs. Whilst it is possible to emulate the entirety of FTT:Power, this would be computationally intensive due to the large numbers of potential model inputs and outputs. We focus on inputs and outputs relevant to the current study, but these could easily be substituted for alternatives within the emulation framework to focus on other parts of the model.

Given the model runs generated by the space-filling design above, we train a Gaussian Process (GP) emulator for each output of interest. A GP is a flexible statistical model that can be used to approximate the true, expensive model, producing efficient predictions of the output at any set of inputs (Couvreux et al. 2021). Importantly, a GP provides uncertainty on these predictions, with interpolation at known points and uncertainty increasing as prediction points move further away (in input space) from observed simulations (Oakley & O'Hagan 2002).

We follow a similar procedure to Couvreux et al. (2021). Denoting by $f(x)$ a particular output given when FTT:Power is simulated at vector of inputs $x$ in parameter space, we assume the general GP form:

$$f(x) | \beta, \sigma^2, \delta \sim \mathrm{GP}\big(m(x, \beta), C(x, x', \sigma^2, \delta)\big).$$

for prior mean function $m(x, \beta)$ with a vector of coefficients, $\beta$, and prior covariance function, $C(x, x', \sigma^2, \delta)$, containing variance parameter, $\sigma^2$, and correlation parameters $\delta$, that control the correlation in each input dimension. Given a set of $n$ simulations $f(x_1), f(x_2) \dots, f(x_n)$, the prior mean and covariance functions are updated conditional on this data, with exact expressions for the resulting predictive mean and variance for any $f(x)$ given via the properties of the multivariate Normal distribution (Williamson et al. 2017), and hence predictions are fast.

There are common choices for each component of the GP emulator, although ultimately their choice is problem-dependent, and should be carefully assessed via validation. Here, we select the mean function by considering various combinations of the inputs, aiming to explain the main relationships in the output (with selection of terms performed similarly to Williamson et al. (2013). Given the mean function, we model the remaining residual variability with a Matern 5/2 covariance, and train the model using RobustGasp (Gu et al. 2019).

Emulators, based on the Gaussian process framework detailed, were trained for capacity of onshore wind and solar PV, and power sector emissions, at the global level for 2030, 2040 and 2050. For India, emulators were built for these outputs as well as the weighted average cost of electricity generation and shares of the energy mix made up by renewables, for the same years. For each, we use 80% of the 500 simulations for training and 20% for testing. Performance is validated using the test data and leave-one-out prediction (Dunne et al. 2022) (see Figure SI.1a-1b). The emulation framework can be extended to higher dimensional output, for example the full time series of the above outputs, by combining GPs with dimension reduction techniques (Higdon et al. 2008; Salter et al. 2019).

### 2.4 Prediction and Analysis

### 2.4.1 Sensitivity analysis

Using the above emulators, we perform 'one-at-a-time' (OAAT) sensitivity analysis (SA) to evaluate the inputs that are the main drivers of variability in the model outputs (McNeall et al. 2024). Each input in turn is varied across its entire range whilst the remaining parameters are kept constant, therefore each SA is performed in a particular subsection of parameter space. For the global analysis, all policy parameters are set to their baseline value of 0 and all other techno-economic parameters to 0.5, corresponding to mean or midpoint values based on the literature and empirical data. For the India analysis, policy parameters are set to 0.5, this is more relevant for the subsequent policy analysis. Whilst there are types of SA that account for sensitivities to multiple inputs simultaneously (McNeall et al. 2024), as a proxy for this we perform OAAT SA with multiple different assumptions about the fixed values (see Figure SI.3-4). The relative sensitivity in outputs is calculated and inputs ranked according to the weighted proportion of variation they drive across all outputs. The results identify the most important drivers of variability and guide the rest of the analyses.

### 2.4.2 Uncertainty Analysis

We assess a range of policy scenarios rapidly by evaluating the emulators at different sets of inputs. By fixing certain inputs at chosen values, within subranges, or according to some probability distribution, we can sample from the input uncertainty under different assumptions and use the emulators to propagate this uncertainty through the model to give an assessment of the output. The emulators allow for the efficient and rigorous comparison and assessment of the robustness of policy scenarios under uncertainty.

For each policy scenario, policy inputs are fixed and 20,000 samples are drawn from the distributions over the unfixed inputs, with the emulator mean and variance evaluated for the relevant outputs and sampled from to account for emulator uncertainty.

We investigate the global power sector emissions in 2030 and 2050 under different levels of global ambition and different assumptions about lead times and cannibalisation rates. Policy scenarios include one at current policy levels, and two with a unified global response at mid and high levels as detailed in Table 2. For the analyses several of the parameters are categorised into different levels e.g. slow and fast lead times, low or high cannibalisation, high and low coal prices etc. These levels represent portions of the parameter ranges which the inputs are varied within. For example, the whole range for coal prices is split in half, and then if the coal price falls into the lower (upper) half, it is categorised as a low (high) coal price. This also applies to electricity demand. Some analyses include aggregates of multiple inputs, like lead times which includes leads for solar, onshore and grid connection. The same process applies but to all inputs jointly i.e. if each of these lead times falls into the lower (upper) half of the range it is categorised as having fast (slow) lead times. This also applies to cannibalisation rates of solar and onshore wind power. Higher resolution analyses could disaggregate these categories.

For India's power sector, we first analyse how different policies perform at achieving capacity levels for 2030. A series of policy packages are designed for India's national power system that feature combinations of instruments at different levels. We predict model output for capacity levels of onshore wind and solar PV at each combination of three levels of each instrument, shown in Table 2, and analyse what proportion exceed 393 GW, the combined capacity projected by the CEA (CEA 2023).

Second, we analyse the interaction between lead times, electricity demand and phase-out instruments by 2050. We compare a mid-level phase-out to current policy.

Finally, we examine a wider range of policies and their robustness in achieving multiple policy objectives under uncertainty. Objectives are represented by chosen target levels for outputs for 2030: 1) the CEA projection of 55% renewables share of generation, but we do not including nuclear and large hydropower, 2) the weighted average cost of electricity generation equal to or below $68/GJ, the 2030 value under current policy derived from the FTT:Power baseline scenario, 3) power sector emissions below 1000 $MtCO_2$/year, representing a reduction of approximately 30% from 2024 levels (Ember 2025). The policies include combinations of all instruments at mid-level (Table 2). We examine the robustness to most important inputs across all outputs by comparing the proportion of emulator predictions that achieve the targets under each policy package subject to variation in the techno-economic parameters. These parameters are chosen based on the SA.

Any probability distribution could be specified across uncertain inputs, with uncertainty in these sampled from and propagated through to the model output via the emulators. Therefore, many different assumptions on inputs can be efficiently compared, and the following results could be reproduced under different starting judgements. For the main analysis we sample uniformly from parameter space (Figure SI.2a) but demonstrate other possible choices (Figure SI.2b).

## 3 Results

We present our results in two parts. Firstly, we explore the key relationships in the model at the global and national level through sensitivity analyses, highlighting the main inputs driving variation, and uncertainty, for energy transition outputs. Secondly, we explore sets of scenarios to compare policy impact and robustness to key uncertainties.

### 3.1 Sensitivity Analyses

At the global level, we focus on the outputs of cumulative capacity of key technologies, solar PV and onshore wind, and power sector emissions. The SA presented in Figure 2 shows the relative importance of inputs in 2030, 2040 and 2050, conditional on no policy and techno-economic parameters set at their median. For contrast, Figure SI.5 shows the same SA but policy variables are set to their maximum, which changes the relative importance of other variables due to how these interact with policy.

Changes in cannibalisation rates for solar and wind drive most of the variation in capacity outcomes. Higher solar cannibalisation substantially increases emissions, while onshore wind cannibalisation has a stronger effect later in time. Cannibalisation has a wide parameter range, however when restricting it to a smaller range (Figure SI.4), the conclusions are similar.

Lead times drive output variation heavily, particularly solar leads. Onshore leads strongly affect its deployment as do discount rates, with impacts diminishing over time. Emissions are sensitive to electricity demand in the near term but then solar cannibalisation dominates.

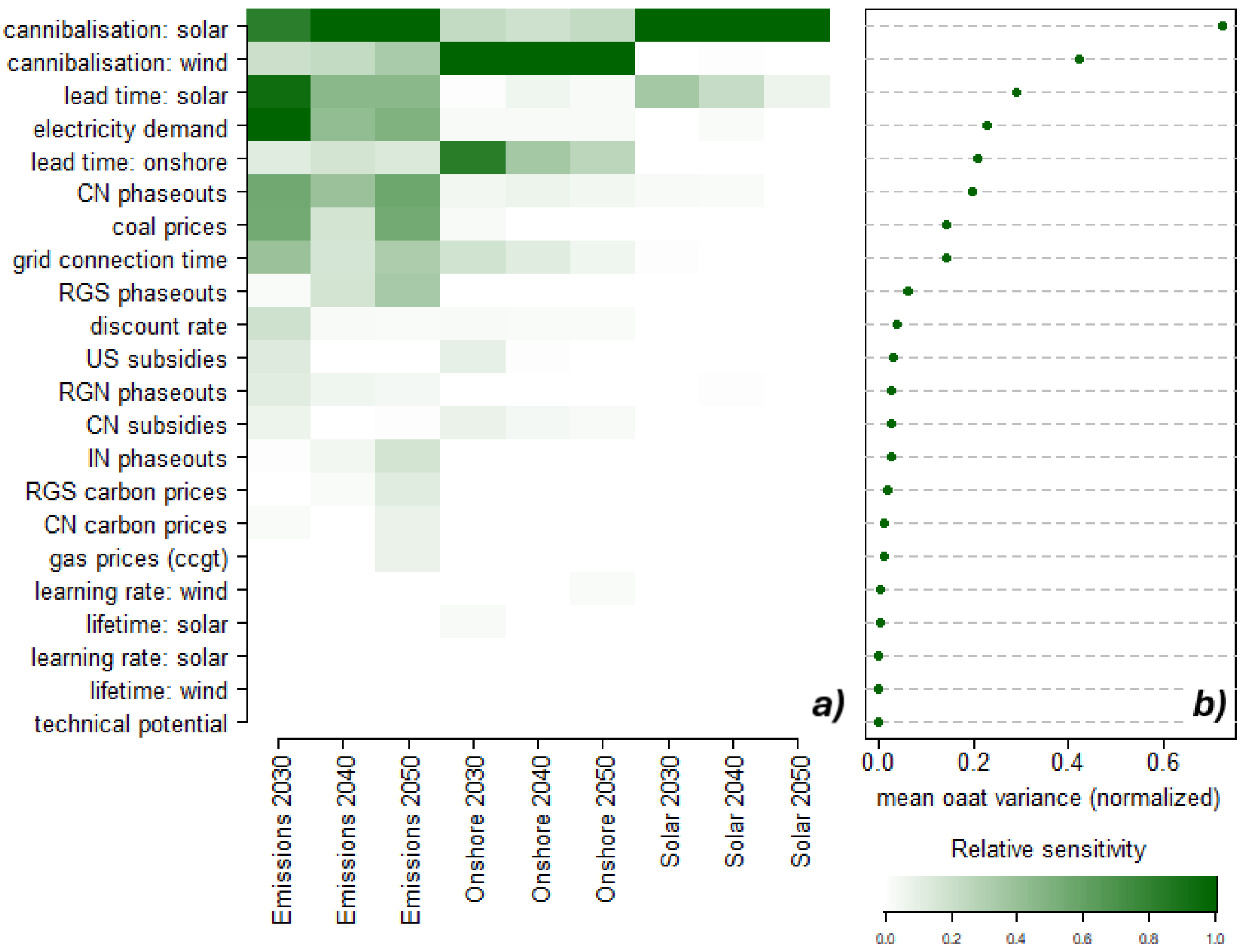

***Figure 2. OAAT sensitivity analysis for global outputs across time**. Panel **a)** shows the relative sensitivity of outputs (x-axis) to policy and techno-economic inputs (y-axis) at three timepoints, policy parameters below 0.001 effect excluded. Panel **b)** shows the average sensitivities across all outputs shown ordered by importance. Abbreviations: China (CN), Rest of Global South (RGS), Rest of Global North (RGN), Solar PV Capacity (Solar), Onshore Wind Capacity (Onshore). Figure generated using code adapted from McNeall et al. (2024).*

In terms of policy inputs, phase-outs have the strongest impact. Phase-out regulations in China substantively affect capacity levels of both technologies but particularly emissions across all years. Phase-outs in India and the rest of the Global South increasingly drive global emissions, while the impact of the Global North declines, reflecting faster electricity demand growth in lower-income regions. Carbon pricing gains in importance in later years for all large regions, partly because it is implemented as a rising cost. Subsidies, and US rollbacks, mostly drive variation earlier through onshore capacity and related emissions saving. Whilst regional policy appears to be less important than techno-economic factors, the global aggregative nature of the latter will bias the results. In reality, techno-economic parameters are highly context-dependent but are implemented this way for tractability of the current analysis and other choices can be made in future analysis.

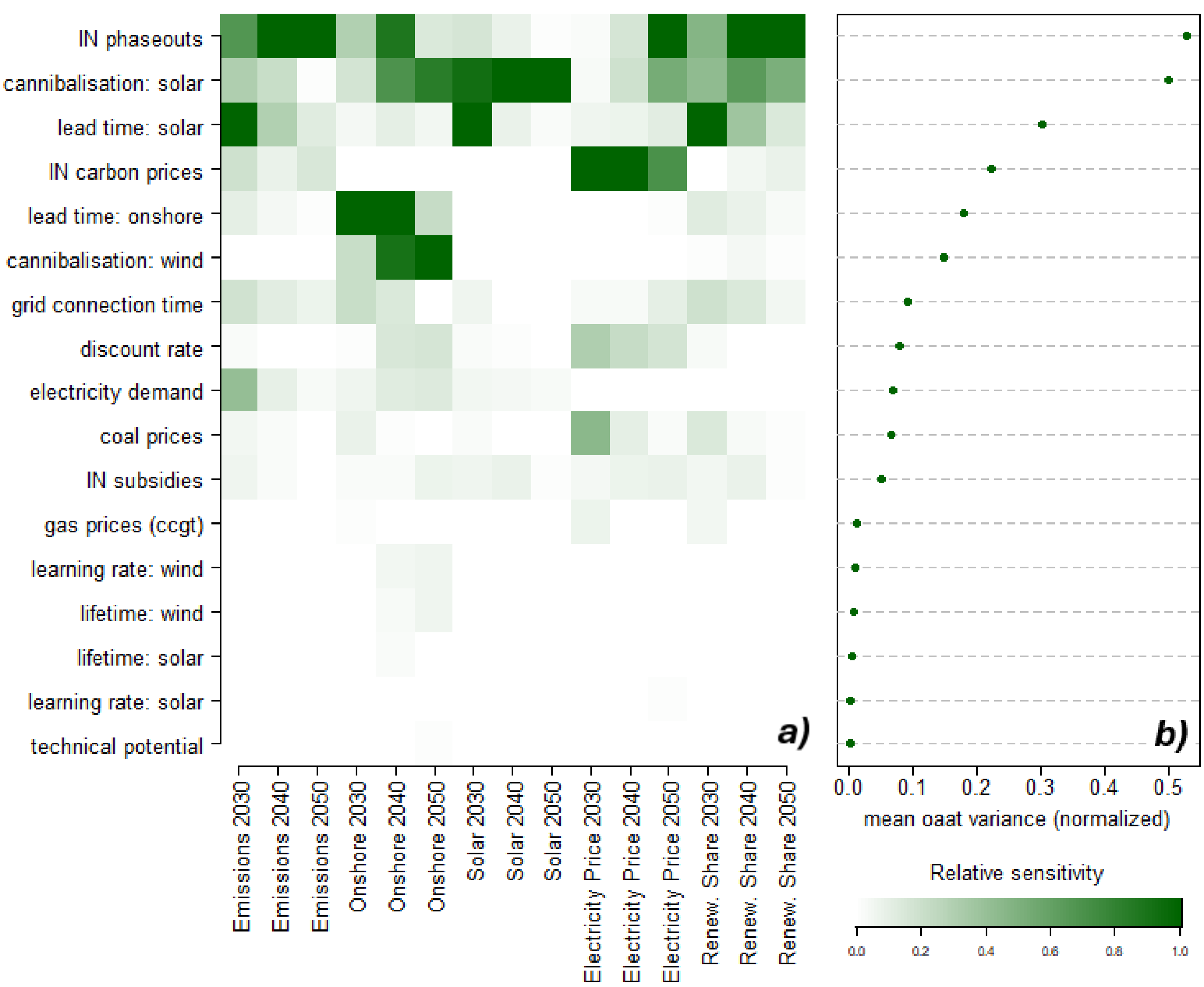

***Figure 3. OAAT sensitivity analysis for India outputs across time***. *Same as Figure 2 for Indian outputs. Electricity price refers to weighted average generation price, Renew. Share refers to share of total capacity provided by renewables excluding nuclear and large hydropower.*

Figure 3 shows the SA for India where all Indian policy inputs are set at 0.5, techno-economic parameters are still at 0.5 (for alternative SA assumptions see Figure SI.5). We see that phase-outs strongly impact onshore capacity and near-term solar capacity. They also increasingly affect emissions and renewables share, but not average electricity prices. Carbon prices have a smaller effect on renewables capacity and emissions and a larger effect on electricity prices more. Subsidies also drive some variation in electricity prices and capacity levels. However, they have little effect on emissions as subsidies mostly switch capacity additions between renewables, rather than replacing fossil fuels.

Capacity outputs are highly sensitive to cannibalisation rates. Solar cannibalisation heavily influences emissions and the share of renewables although this impact weakens over time. Coal prices influence both the rate of rollout of renewables as well as the electricity price, influenced by the remaining coal in the power mix. Growth rates of electricity demand and lead times are important, particularly in the near term.

The results of the above SA provide insight into what inputs drive most variation, and therefore uncertainty, in outputs of interest. This informs which inputs feature more prominently in subsequent analyses.

## 3.2 Uncertainty Analyses

### 3.2.1 Global Power Sector Emissions Over Time

Figure 4 shows how lead times, cannibalisation rates and different levels of global policy ambition interact to influence the level of emissions in 2030 and 2050, under uncertainty in all other techno-economic inputs.

Stronger policy ambition lowers median emissions in all scenarios and makes long-term outcomes more constrained, as shown by the narrower distributions. Scenarios with slow lead times tend to have higher emissions but are more constrained in the short term. In contrast, scenarios with fast lead times and high ambition are more predictable in the long term. Fast lead times lead to much larger emissions reductions, even under baseline policy. Assuming high cannibalisation reduces these gains but does not remove them.

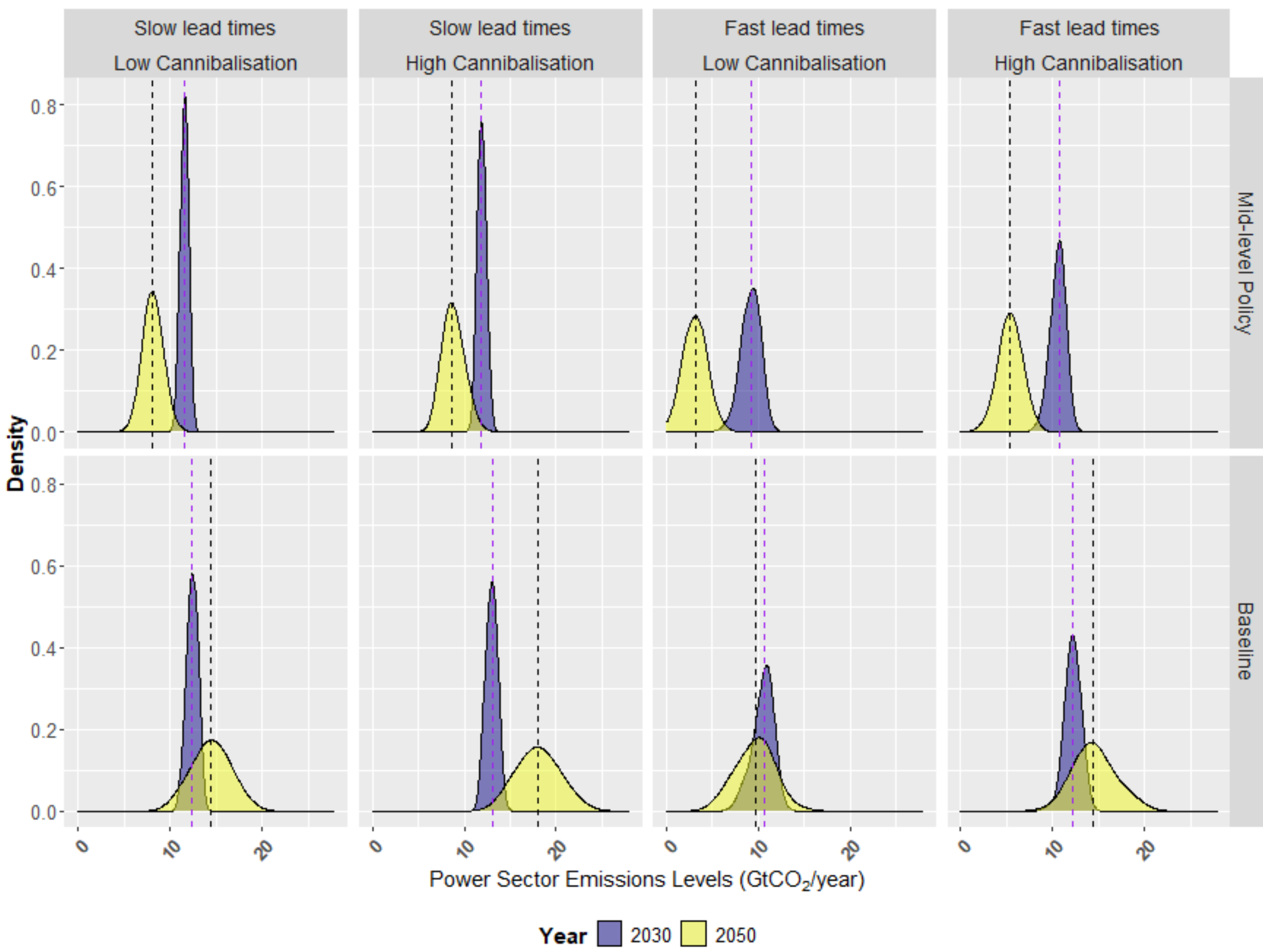

***Figure 4. Power sector emissions distributions under different policy-techno-economic scenarios**. Distributions represent the variation due to the other uncertain techno-economic inputs within that scenario. Dotted lines are median levels. Lead times are compared with the next tile but one in the same row, e.g. for the baseline observe the bottom left tile, and compare with third tile along. Cannibalisation rates are compared with the other tile in the same half of the plot. Policy impact is compared with tiles directly above or below.*

High cannibalisation results in higher median emissions for a given lead time and longer lead times lead to higher emissions for a given cannibalisation rate. Absolute policy impacts are larger in high-cannibalisation scenarios but result in higher emissions. For example, with fast leads and high cannibalisation (the right-most column), median emissions fall from about 14.5 to 5.5 $GtCO_2$/year under mid-level policy, a reduction of 9$GtCO_2$/year. With slow leads and high cannibalisation (second column), the absolute reduction is larger at 9.3$GtCO_2$/year. However, assuming slow leads and low cannibalisation (first column), the emissions reduction is only 6.3$GtCO_2$/year.

Figure 4 shows how the impact of different uncertainties changes with policy strength. Under baseline policy, by 2050 the relative impact of higher cannibalisation results in a median increase of between 3.6-4.7$GtCO_2$/year, depending on lead times. Under mid-level policy, higher cannibalisation increases emissions by only 0.6-2.3$GtCO_2$/year.

Policy ambition also strongly affects the shape of the distributions, the $5^{th}$-$95^{th}$ percentile range of emission levels are 33% smaller than in the baseline scenarios which spans 12.2 $GtCO_2$/year by 2050, see Figure SI.6. Stronger policy leads to both lower emissions and reduced variability under uncertainty in FTT: mid-level policy ensures that emissions will be lower in 2050 than 2030 whereas in the baseline scenarios greater emissions in 2050 is the most common outcome. This analysis features highly aggregated assumptions about global policy that should inform interpretation. For example, the 3 policy scenarios assume universal policy action and uniform levels of policy stringency for the 3 policy parameters. We feature a more nuanced treatment of policy strategies in the Indian analysis.

3.2.2 India Policy Instrument Comparison for Capacity

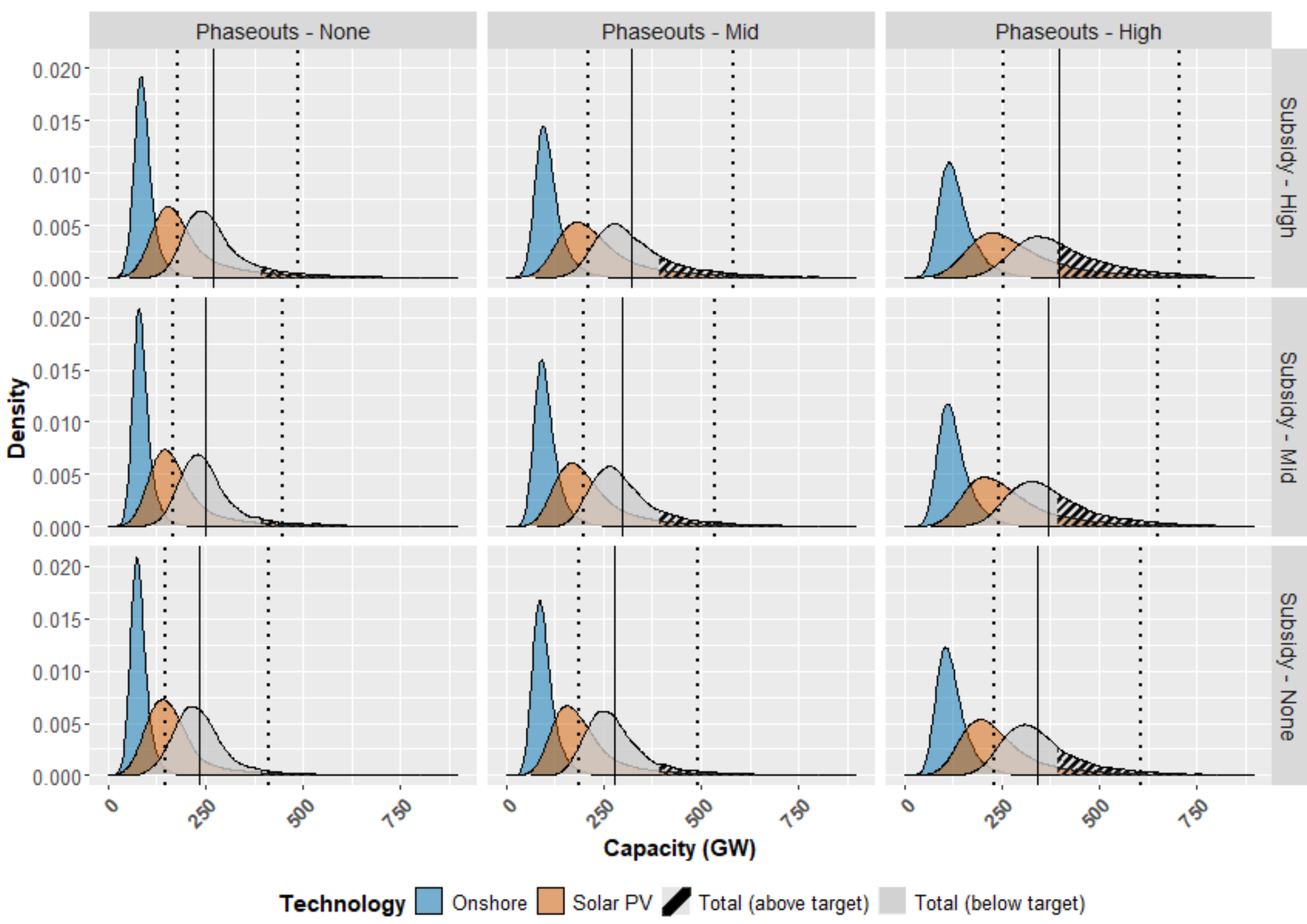

***Figure 5: Distributions for Indian renewable energy capacities in 2030 across national policy packages.*** *Panels represent different policy instrument combinations. Each distribution represents the capacity of onshore, solar PV and their total under full variation in techno-economic inputs across their ranges. The total has 90% interval range marked by dotted lines with black solid lines representing the median total capacity. Hatched sections represent predictions above 393 GW. Subsidy levels are compared vertically. Phase-outs are compared horizontally. The plots are truncated at 800 GW, though a few samples exceed 1000 GW under high policy.*

Figure 5 shows solar PV and onshore wind uptake in India in 2030 under different combinations of phase-outs and subsidies (see Figure SI.7 for a comparison between phase-outs and carbon pricing). Fossil fuel phase-outs strongly drive uptake of solar and onshore capacity and result in large interactions with favourable conditions, widening the distributions and skewing them towards higher values. The 95$^{th}$ percentiles (right dotted lines) rise more than the medians as the phase-out stringency increases, showing that phase-outs interact with uncertain inputs, such as lead times or cannibalisation rates, where favourable assumptions about these could lead to more rapid uptake. Even without policy solar distributions are wider and more skewed than onshore due to sensitivity to techno-economic uncertainty, showing that solar can grow rapidly under favourable conditions. Subsidies have a stronger effect when combined with some level of phase-out, as shown by the larger increase in median capacity under higher phase-out levels. Subsidies make renewables more attractive relative to other technologies, particularly when coal generation is partially restricted through phase-outs.

The distributions reflect scenario outcomes rather than probabilities; they indicate conditions under which targets may be achieved. Under baseline policies or mid-level phase-outs, few scenarios meet the target, even given optimistic values of uncertain inputs. Fast lead times, low cannibalisation, high coal prices, and high electricity demand substantially increase the portion of scenarios exceeding it especially under high policy. These results show not only the impact of policy but also the variability of outcomes under any policy scenario when subjected to a comprehensive treatment of plausible variability, and therefore uncertainty, in the techno-economic context. This has substantial implications from a planning perspective and in the next subsection we perform more systematic exploration of these uncertainties for emissions.

3.2.2 Phase-out Robustness for Emissions Reduction

Figure 6 shows Indian power sector $CO_2$ emissions as a function of phase-out levels (the most important instrument, according to the SA), and the most influential techno-economic parameters: coal prices, cannibalisation rates and electricity demand. Even partial phase-outs substantially reduce median emissions. Without any phase-out policy, emission levels are more influenced by variation in demand growth rates. In 2050 under current policy, model output for emissions could be anywhere between 0-5500 $MtCO_2$/year. With a partial phase-out this range is restricted to 0-3000$MtCO_2$/year.

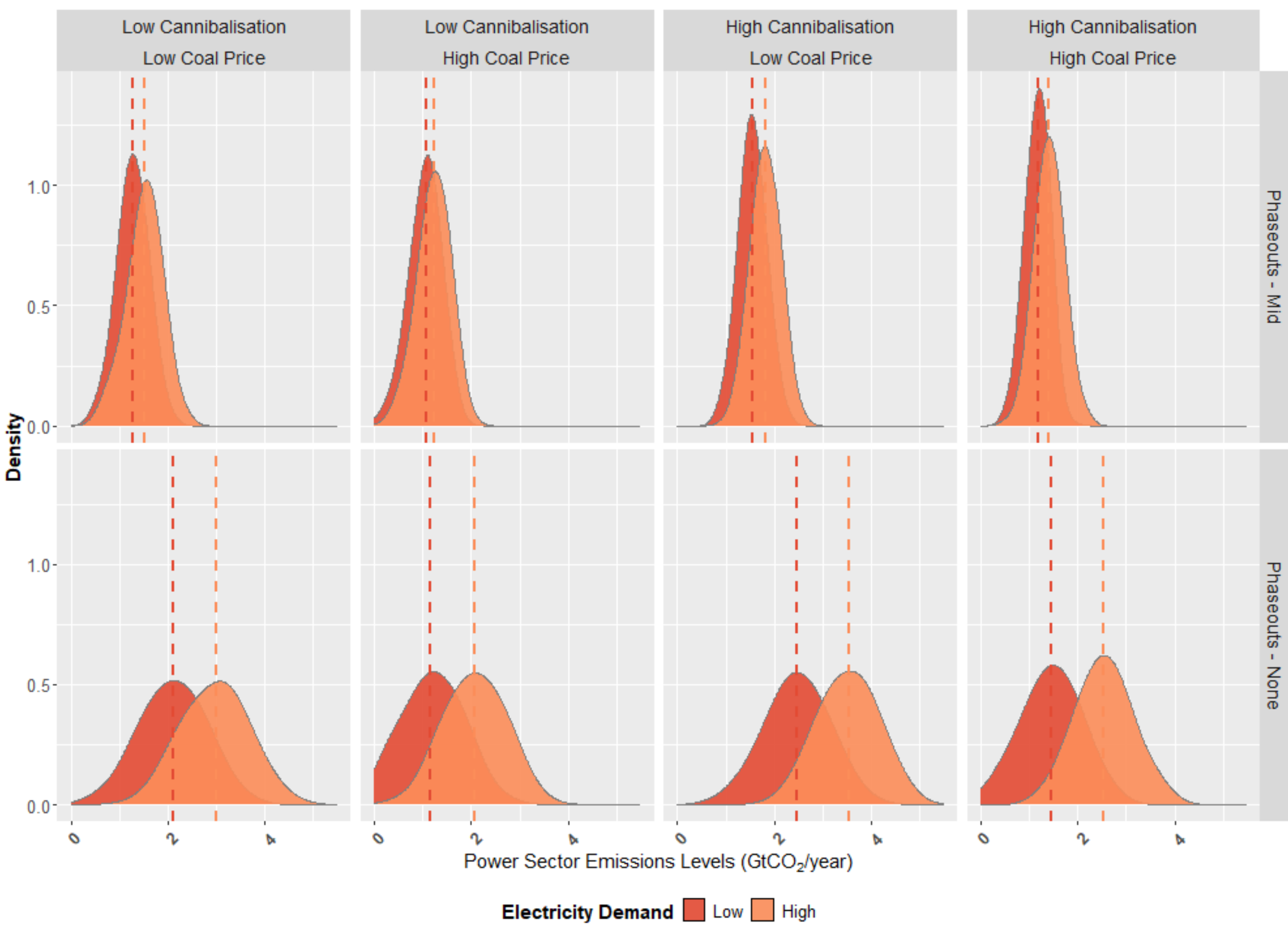

***Figure 6: Distributions of Indian power sector emissions under policy and electricity demand variation.*** *Rows represent different levels of phase-outs, columns represent different combinations of cannibalisation rates and coal prices. The coloured curves represent different years; the shades represent levels of electricity demand. Dotted coloured lines represent the median for each demand level.*

The ranges of the distributions under no additional policy are substantially decreased under a partial phase-out scenario. The shape of the distributions in the scenarios without phase-outs remain consistent, even if demand shifts median levels strongly. Under low cannibalisation, low coal prices increase annual emissions by 900Mt$CO_2$, relative to scenarios with high coal prices regardless of whether demand is high or low. Variation in demand for a given level of cannibalisation has similar impacts on median emissions.

Under partial phase-outs the variability of emissions is reduced, with reduced sensitivity to electricity demand as shown by the overlap of distributions and less change in median levels particularly under low cannibalisation. Tall and thin distributions reflect the added constraint on emissions in mid-policy scenarios.

### 3.2.3 Policy Robustness Over Multiple Policy Goals

Our results from the two analyses above show phase-outs having substantial impact on the transition in India for individual policy goals. Here we examine a broader set of policies and goals for India in 2030. Figure 7 displays the proportions of emulator predictions that achieve target levels for emissions, renewables share and electricity prices.

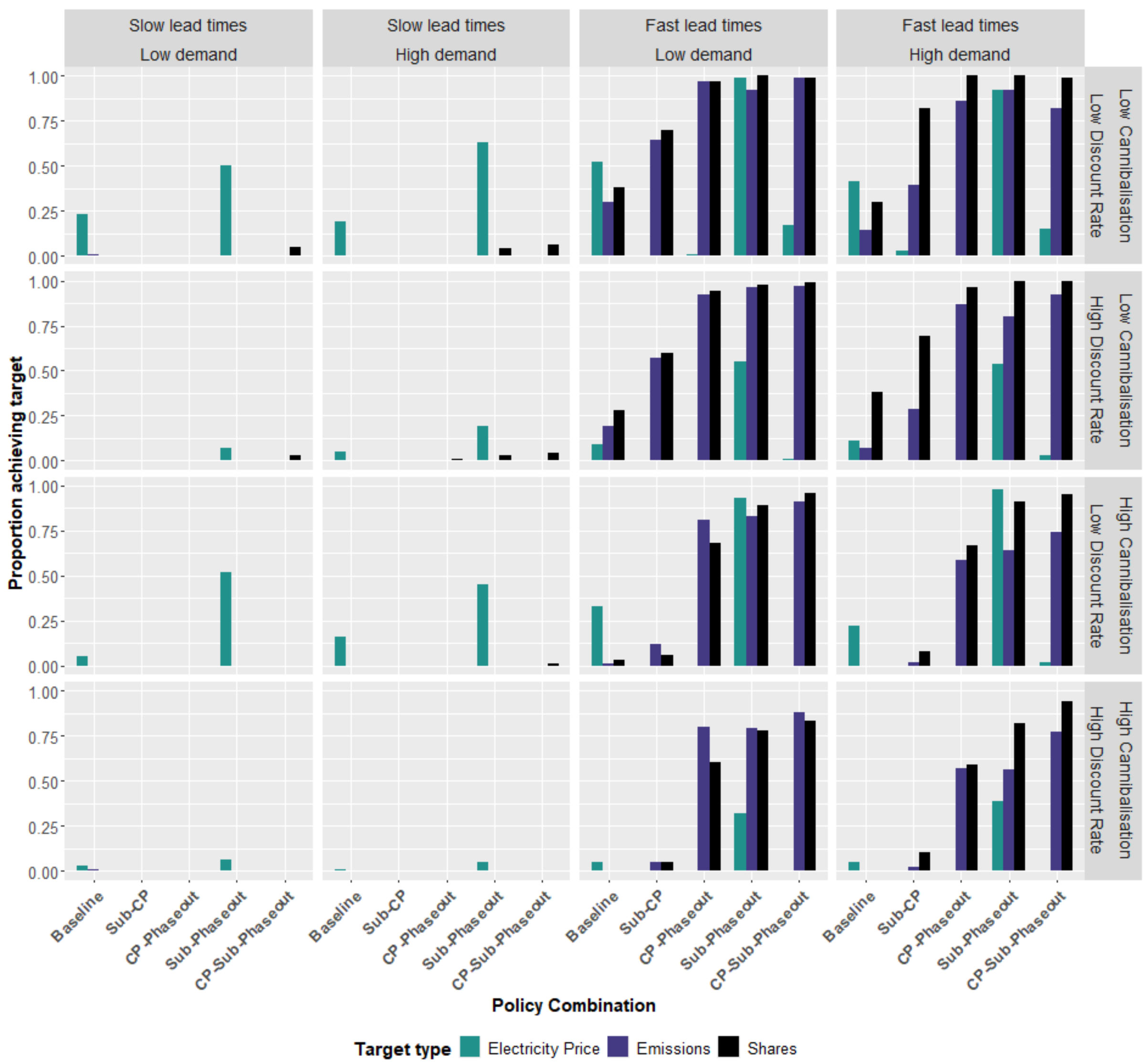

***Figure 7. Proportions of predictions meeting 2030 targets in India.*** *Panels represent different combinations of ranges of techno-economic parameters. Lead times, demand, cannibalisation and discount rates are varied uniformly within their subranges, identified by the panel, and all other techno-economic parameters are varied uniformly across their whole parameter ranges. Bars represent the proportion of samples that achieve targets. Bars are separated into different policy packages: baseline, subsidies-carbon price (Sub-CP), carbon price-phase-outs (CP-Phase), subsidies-carbon price-phase-outs (Sub-CP-Phase), subsidies-phase-outs (Sub-Phase).*

Policy goals targeting renewables share or emissions levels are rarely achieved in any of the slow lead time scenarios and electricity price targets are reached under low discount rates only.

When lead times are assumed to be fast, even baseline policy achieves all policy goals in a small proportion of scenarios when cannibalisation remains low. A purely market-based policy, using subsidies and carbon pricing, attains emissions and market-share targets only within a limited region of the scenario space, exceeding 60% of runs only when demand, discount rates and cannibalisation are low. Furthermore, emissions reductions remain sensitive to demand, although higher demand benefits capacity development. High cannibalisation rates severely reduce the number of samples achieving any target.

Assuming fast leads, policy packages that include a partial phase-out are much more robust to the other uncertainties. High cannibalisation substantially reduces performance of all packages. The inclusion of carbon pricing means electricity prices rise above the target in all samples. The subsidy-phase-out package achieves the price goal most often but remains sensitive to discount rates due to front-loaded cost profiles of renewables. Coal price is omitted from Figure 7 but is important for all policy goals and has similar impacts to discount rates, see Figure SI.8.

Lead times are the most impactful input, and all policies have limited robustness to them. Phase-out based instruments exhibit strong robustness to electricity demand, cannibalisation and discount rate uncertainty for energy share and emissions goals, in comparison to market-based only instruments. When phase-outs are coupled with a subsidies, the package is more robust to all uncertainties, including lead times, in achieving the electricity price target. In conclusion, some packages are more robust to specific uncertainties in the model and the robustness of all policies would be increased if accompanied by policies that decrease lead times, restrain cannibalisation and lower the costs of finance.

## 4. Discussion & Conclusion

Our methodology enables new treatments of uncertainty that are typically too computationally costly to perform using simulations alone. Using emulators we replicate established methods, incorporate in-depth uncertainty analysis and apply this to policy design. We demonstrate a general methodology for uncertainty analysis in modelling research and apply it to the power system transition model, FTT:Power, with a focus on policy robustness. In this application, we adopt flexible assumptions about policies, regions and general input parameterising to explore the energy transition globally and in India.

Our analysis addresses parametric uncertainty only, and thus it reflects the structural errors of the model (Pastor et al. 2020). For example, FTT:Power previously utilised generalised LCOE comparisons of technologies which is criticized for misrepresenting what motivates investors, with profit proposed as the key decision driver (Ehrenmann & Smeers 2011). We partially address these concerns by including a representation of cannibalisation making the comparisons based on a value-adjusted LCOE, but the implementation could still bias results, making renewables more attractive investments in certain contexts. We also treat cannibalisation as a function of penetration alone which may fail to capture how policy choice affects cannibalisation rates, with some evidence that carbon taxes lead to less cannibalisation than subsidies (Brown & Reichenberg 2021; Liebensteiner & Naumann 2022).

Furthermore, elements of our approach, such as using global averages and assigning wide parameter ranges to indeterminate inputs, results in other biases. Output ranges will be wide and this could overstate the benefits of stricter policy from a robustness perspective. These structural elements frame the policy implications of the study and we present possible remedies and future research directions below. However, we argue that the implementation choices are justifiable and the scenarios included are illustrative. Modelling attempts at explicitly incorporating cannibalisation dynamics in global power system models which integrate into a macroeconomic frameworks remain rare (Gong et al. 2023). We do so with wide uncertainty ranges that capture several proposed rates across the literature. Whilst any structural biases are naturally replicated by the emulator, the approach allows the study to be replicated easily with different assumptions over the inputs. It can also be applied to different model structures, and emulators are commonly used in calibration problems to explore and address structural biases in climate models (Prévost et al. 2026; Williamson et al. 2015).

We find that uncertainty in cannibalisation rates of key renewables, plant construction and grid connection speed (lead times) drives much of the uncertainty in a global transition. Cannibalisation threatens renewables deployment, yet empirical research on possible rates remains limited. Longer lead times also jeopardise policy goals, and interacting uncertainties widen the range of model outcomes, though policy can both mitigate emissions and narrow this range. Solar, despite large cost declines and policy support in recent decades, is still sensitive to lead times and cannibalisation, but resilient to other factors. Onshore wind is affected by additional factors, including demand, discount rates and fossil fuel prices. For India, our results show that in FTT, pure market-based approaches are effective in a narrow set of circumstances, but miss all policy targets in most scenarios. In contrast, partial fossil fuel phase-outs coupled with carbon pricing or subsidy support for renewables or both showed robustness against a wide range of uncertainties, but remain sensitive to lead times. We infer that policy efforts targeting the root causes of cannibalisation and construction delays, combined with regulation constraining fossil fuel deployment enable quicker and more robust transitions in the power sector.

We investigate highly uncertain cannibalisation through keeping assumptions weak, exploring a wide range of values covering most estimates suggested in the existing literature and exploring alternative assumptions, see Figure SI.4. In our analysis, faster cannibalisation can severely restrain the diffusion of renewables, especially if it outpaces experience-based cost reductions and is not tackled through policy. Our analysis in India suggests that regulation-based policy-packages are robust to higher levels of cannibalisation whilst market-based packages are not.

Within FTT:Power, longer lead times delay revenue creation, which limits the speed of reinvestment into the next generation of power plants. In other words, the longer lead time impacts the maximum growth rate. Technologies with fast lead times can most rapidly fill in market gaps compared to technologies with longer lead times (Mercure 2015). Given the strong path dependency in the model, with growth proportional to existing shares, this has long-term impacts (Mercure et al. 2014). In earlier stages of the transition, lead times for renewables were relatively low. These have been increasing over time for most renewables, but solar PV remains the quickest to commission (Gumber et al. 2024). Coupled with our findings, this suggests that lead times are a new key and growing driver of uncertainty in the global transition.

Regional energy policies, both supportive and adversarial, now drive relatively less uncertainty at the global level for the expansion of solar PV. This supports previous work highlighting the momentum of solar expansion (Nijsse et al. 2023) and this limits the impact of regional rollbacks, for example, the US's removal of aspects of the Inflation Reduction Act. Under our assumptions about plausible parameter uncertainty, direct policy uncertainty (prices and regulation) in individual states has a limited relative impact on the diffusion of solar PV, although Chinese policy impact is still noticeable. Onshore wind is also impacted by China, but the US impact is restricted to the near term.

However, this does not include all policy levers and does not capture how the US might exert political pressure on other countries (MacNeil & Paterson 2020), or the US's role in international financial institutions (Merling & Forster 2024; Helleiner 2025). Further research is needed into policy affecting enabling factors to give insight into how regional policy exerts influence and drives uncertainty but our analysis supports that solar's diffusion is resilient to some policy uncertainties. Policy uncertainty affects onshore deployment more strongly. In both the global and the regional analysis, phase-outs and subsidies have strong impacts on deployment. Without such support, onshore rollout is more sensitive and in certain scenarios its diffusion could be jeopardised by both competition from fossil fuels and solar, which could compromise grid stability given the complementary load profiles of solar and wind.

In comparison to other benchmark projections such as that of the IEA, we include a broader set of uncertainties which makes the case for strong regulatory policy more compelling. The inclusion of more uncertainties results in much higher sensitivity than the IEA's analysis, underscoring our contribution. Our findings suggest policies targeting the drivers of cannibalisation, decreasing lead times and enabling financing are critical. Policy that targets storage capacity, grid capacity and flexibility mechanisms such as interconnectors and demand-side responses could constrain cannibalisation (Stiewe et al. 2025; López Prol et al. 2020; IEA 2024). More fundamental changes in energy system and market design could be necessary to mitigate such long term risks (Gonzalez-Ruiz et al. 2026). These range from greater use of Contracts for Difference for hedging (Beiter et al. 2024) to greater hybridisation of markets (Dimanchev et al. 2024) and public ownership of power systems (Christophers 2024). These debates reach far back in the economic literature (Weitzman 1974) but such discussions are beyond our scope here. Our contribution highlights the scale of the potential acceleration of policy impact on power sector transitions if progress is made on tackling these uncertainties globally, either through addressing root causes or creating enabling external system or market conditions.

India is a growing player in global climate change mitigation. Whilst previous projections fall within our range of outcomes (Thambi et al. 2018; Das et al. 2023), our findings show how plausible variability in techno-economic parameters can lead to substantial uncertainty in those estimates. Our median results suggest that the previous projections regarding the share of renewables are conditional on favourable lead times and cannibalisation rates. This is a strong assumption and these studies often do not include a comprehensive assessment of parametric uncertainty beyond a few key inputs (Laha & Chakraborty 2021; Fern Lobo et al. 2023). Previous scenario modelling performed for India often produces widely differing results without detailing assumptions (Spencer & Dubash 2022). Our results support that the large uptake of solar and other renewables is already displacing coal at an accelerating pace (Joshi et al. 2025; Das et al. 2023). However, the pace of the transition remains uncertain and plausible bottle necks could undermine its momentum, and conventional modelling methods struggle to assess how uncertainties endanger policy targets. Our approach makes these uncertainties more transparent and enables clearer evaluation of how different policies perform against policy goals.

We show that subsidy-phasedown packages show significant robustness to many uncertainties, but slow lead times still jeopardise policy targets. This aligns with previous research citing the expansion of supporting infrastructure as a key obstacle (Debanjan & Karuna 2022; Vijayalaxmi & Manthanwar 2024). The expansion of solar PV is now resilient to some uncertainties but if this is not accompanied with policies supporting other renewables this could slow uptake and endanger grid stability (Debanjan & Karuna 2022). Our results suggest that policy packages that do not include some form of regulations on fossil fuels are not likely to be robust. However, the configuration of subsidies and taxes is also important and both instruments have the potential to progress policy goals although they do this in different ways and also differ in how costs are distributed. Furthermore, the justice implications of phase-outs in this context should also be considered thoroughly (Pandey & Kumar 2025; Vishwanathan et al. 2025).

Our application could be easily implemented with different assumptions, parameter mappings and objects of interest etc., to investigate different aspects of parametric uncertainty within the FTT framework. Structural uncertainties could be addressed by application of the methodology to other power system models for comparison. Most easily, as FTT undergoes development, different model versions could be compared using our workflow for insight in how such developments affect conclusions about uncertainty and policy robustness (Few et al. 2023). Our results could be compared with the mathematical programming approaches mentioned throughout this paper which account for dynamics such as profit-maximisation, market power or long-term risk aversion. This could provide crucial complimentary insights to our findings (Devine & Siddiqui 2023; Pineda et al. 2018). Calibration approaches using emulators, such as history matching, are commonly used in other fields (Williamson et al. 2013; Salter & Williamson 2016; Couvreux et al. 2021). These could be readily integrated into our approach for validation studies, insight into model discrepancy and deeper uncertainties. Furthermore, our methodology could be applied to different models in different policy domains. However, the use of any modelling approach implies particular underlying framings (Stirling 2023) which a comprehensive treatment of uncertainty would consider (McLaren & Markusson 2020; Brugnach & Ingram 2012).

Our methodology enhances understanding of uncertainties conditional on the structural assumptions embedded in chosen model framework (Stirling 2010). Comparing our insights with those captured from approaches under different assumptions is crucial for contextualising our findings. Nevertheless, we argue that our methodology provides a more extensive treatment of uncertainty than is seen in much climate policy modelling research and it could be extended to explore deeper sources of uncertainty and conditionality. We have shown that comprehensively accounting for uncertainties in model inputs has important implications for modelling results and policy performance. Different assumptions lead to dramatically different results, and this becomes especially important when considering policy targets. In view of this, we highlight priorities for policy targeting power system transitions, particularly regarding cannibalisation and lead times for renewable generation technologies as well as the importance of phase-outs at both the global and Indian national level. Our approach offers a novel way of analysing uncertainty for climate policy and evaluating the policy robustness to those uncertainties.

**Acknowledgements**

For the purpose of open access, the author has applied a Creative Commons Attribution (CC BY) licence to any Author Accepted Manuscript version arising from this submission. This project was supported by a UKRI Centre for Doctoral Training in Environmental Intelligence PhD studentship hosted by the University of Exeter (EP/S022074/1).

**Supplementary Information**

Historical generation and capacity of renewable energy from IRENA is available at https://www.irena.org/Publications/2025/Mar/Renewable-capacity-statistics-2025. Original data from BNEF and IEA are licensed by these owners, but datasets derived by the authors are available as part of the model code (see repository).

Code and data used for the analysis can be found in the following repository: https://github.com/cpmodel/FTT_StandAlone/tree/data_update_main_emulation